\documentclass{article}
\usepackage{spconf,amsmath,graphicx,multirow,booktabs,pifont,comment,amssymb,enumitem}
\usepackage[table,xcdraw]{xcolor}

\title{Exploration of Adapter for Noise Robust Automatic Speech Recognition}
\name{Hao~Shi,~Tatsuya~Kawahara}
\address{Graduate School of Informatics, Kyoto University, Kyoto, Japan}
\begin{document}
	\ninept
	\maketitle
	\begin{abstract}
		Adapting an automatic speech recognition (ASR) system to unseen noise environments is crucial. 
Integrating adapters into neural networks has emerged as a potent technique for transfer learning. 
This study thoroughly investigates adapter-based ASR adaptation in noisy environments. 
We conducted experiments using the CHiME--4 dataset. 
The results show that inserting the adapter in the shallow layer yields superior effectiveness, and there is no significant difference between adapting solely within the shallow layer and adapting across all layers. 
The simulated data helps the system to improve its performance under real noise conditions. 
Nonetheless, when the amount of data is the same, the real data is more effective than the simulated data. 
Multi-condition training is still useful for adapter training. 
Furthermore, integrating adapters into speech enhancement-based ASR systems yields substantial improvements. 
	\end{abstract}
	\section{Introduction}
	\label{sec:intro}
	\vspace{-5pt}
Automatic speech recognition (ASR) in noisy environments is still challenging \cite{6732927}. 
Deep learning-based ASR models \cite{8462506, 10.1007/978-3-319-22482-4_11}, however proficient, exhibit susceptibility to performance deterioration when exposed to unencountered noise sources. 
In today's large-scale models \cite{NEURIPS2020_92d1e1eb, 9585401}, the imperative to adapt models to new noisy conditions with minimal data becomes crucial. 
Domain adaptation \cite{9864219} in ASR involves finetuning a model from one (source) domain to work effectively in another (target) domain, managing challenges arising from distribution disparities. 
Data augmentation \cite{7113823, 9689650}, transfer learning \cite{9054543}, domain adversarial training \cite{8462663}, and multi-task learning \cite{10096615} have shown beneficial for adapting the model to specific tasks.

	Adapter, which belongs to transfer learning \cite{9746223}, has been very popular recently. 
	It has been applied to various ASR tasks, empowering models to adapt efficiently to specific challenges. 
	Adapters serve as a crucial bridge in addressing various low-resource ASR tasks \cite{9864219,9855860,pmlr-v202-yu23l,10094903,9746126,10023323}. 
	Moreover, they have shown effectiveness in the disorder and children ASR \cite{9864219,9746126,10023323}. 
	Furthermore, when applying the pretrained model \cite{NEURIPS2020_92d1e1eb, 9585401}, how to better transfer the knowledge in a large model to new scenarios with limited data is particularly important. 
	Because the adapter meets such requirements, inserting the adapter into a pretrained large model has also been widely studied \cite{9864219,9746223,10094903,10095392}.

	Despite the widespread application of adapters across various ASR tasks, limited investigation has been conducted towards noise environment adaptation for ASR. 
	In this paper, we address this problem. 
	Our primary focus encompasses investigating the adapter insertion points, the data employed for adapter training, and the synergy with speech enhancement (SE) front-end models.

	\section{Preliminaries}
	\vspace{-5pt}
	\subsection{Conformer-based ASR System}
	\vspace{-5pt}
	Conformer-based ASR systems \cite{Gulati2020} have demonstrated leading-edge performance across various benchmark datasets. 
	Typically, these systems comprise two core components: an encoder and a decoder. 
	The encoder processes the input speech signal and generates a sequence of feature vectors, encapsulating pertinent linguistic information. 
	Multiple Conformer layers systematically process the input sequence, each operating at varying abstraction levels. 
	Meanwhile, the decoder is tasked with transforming the feature vector sequence, generated by the encoder, into a character or phone sequence representing the transcription of the input speech.

	\vspace{-5pt}

	\subsection{Speech Enhancement-based Robust ASR}
	\vspace{-5pt}
	Robust ASR systems \cite{10.1007/978-3-319-22482-4_11,6288816,9103053} are purpose-built for challenging acoustic environments. 
	This paper narrows its focus to addressing background noise. 
	There are two primary ways to enhance ASR robustness. 
	The first approach, multi-conditioned ASR, involves training the ASR system with diverse noisy datasets \cite{10.1007/978-3-319-22482-4_11}. 
	However, while effective, this method's performance tends to be limited in exceedingly noisy or low signal-to-noise ratio (SNR) scenarios. 
	The second approach involves incorporating an SE front-end \cite{7177943}. 
	This requires separate pretraining of the SE front-end and the ASR back-end, often followed by adopting joint training \cite{9383615}. 
	The joint training loss function combines the ASR loss ($L_{ASR}$) and the SE loss ($\mathcal{L}_{SE}$). 
	It should be noted that computing SE loss ($\mathcal{L}_{SE}$) requires the original clean speech, making it infeasible for real-world noisy data scenarios. 
	As a result, during joint training, only the ASR loss ($L_{ASR}$) is utilized. 
	
	\vspace{-5pt}

	\subsection{Adapter}
\begin{figure}[h] 
		\centering 
		\includegraphics[width=0.4\textwidth]{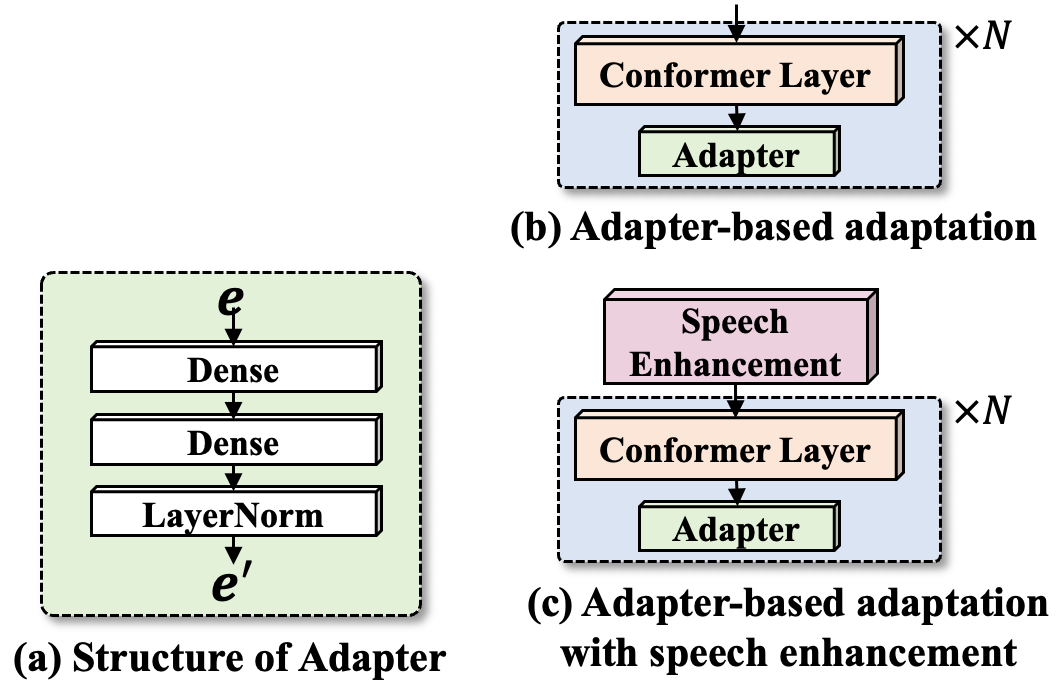} 
		\vspace{-10pt}
		\caption{(a) Structure of adapter; (b) flowchart of the adapter-based adaptation; (c) adapter-based adaptation with speech enhancement front-end. } 
		\vspace{-10pt}
		\label{flowchart} 
	\end{figure}

	\vspace{-5pt}
	The adapter was initially proposed \cite{NIPS2017_e7b24b11} to enhance transfer learning in natural language processing tasks, particularly when faced with limited task-specific data. 
	Its strength is task customization without significantly altering the model's structure. 
	This versatility makes it suitable for various contexts, including accent recognition, emotion detection, and noise-robust ASR. 
	The design of an adapter depends on task requirements and the model architecture \cite{9855860}. 
	The simple yet effective neural network structure of an adapter consists of two dense layers. 
	It takes input from a chosen layer in the pretrained model's output. 
	The primary function of the first dense layer is to capture the initial transformations of input features from the pretrained model. 
	The second layer builds on the transformed features from the first dense layer and captures more intricate task-specific patterns and representations. 
	The adaptation process is depicted as follows: 
	\begin{equation}
		\setlength{\abovedisplayskip}{0pt}
		\setlength{\belowdisplayskip}{0pt}
		e' = e + \text{adapter}(e)
		\label{eq1}
	\end{equation}
	Here, $e$ represents the selected layer's output, and $e'$ represents the adapted feature. 
	The structure of the adapter is shown in Figure~\ref{flowchart}--(a). 
    We employ the LoRA \cite{hu2021lora} structure wherein the central dimension is smaller than both the input and output dimensions.

	\section{Exploration of Adapter for \\ Noise Robust ASR}
	\vspace{-5pt}
	Incorporating adapters into ASR models is a prevalent practice for tasks like accent ASR, children ASR, and multi-lingual ASR. 
	However, noise-robust ASR, another application that necessitates adaptation, has received limited attention in adapter studies. 
	Consequently, this paper delves into the application of adapters for enhancing noise-robust speech recognition. 
	In order to gain a deeper understanding of the adapter's role within this context, our study takes a comprehensive approach, thoroughly investigating its effect from multiple perspectives. 
	\begin{itemize}[itemsep=0pt,topsep=2pt,parsep=2pt, leftmargin=12pt]
		\item[\ding{111}] Where should the adapter be inserted? \\
		The position where the adapter is inserted notably impacts the model's performance. 
		To address this, some methods employ machine learning techniques to automatically identify the optimal layers for adapter integration \cite{10095392}. 
		However, it is worth noting that the best place to insert the adapter can differ based on the specific task. 
		To gain a clear and intuitive understanding of the adapter's influence on noise-robust ASR, we investigate the most effective layer for adapter insertion and assess whether stacking adapters yield improved effect.

		\item[\ding{111}] How to configure the embedding nodes in the adapter? \\
		Modifications to the network structure can have a substantial impact on model performance. 
		However, in existing adapter-based research, focus is often limited to a single chosen embedding dimension. 
		Therefore, the impact of the embedding dimension in the adapter on noise-robust speech recognition is explored.

		\item[\ding{111}] How training data affects the adapter? \\
		Training data is crucial for deep learning-based ASR. 
		To explore the effects of different data quantities and types on adapter-based noise-robust ASR, we conducted experiments employing diverse training datasets. 
		We evaluate how simulation and real training data with varying data quantities affect adaptation. 
		Furthermore, we investigate whether incorporating simulated data could enhance the model's performance on real data. 
		Moreover, we contrast the impact of models trained on a single noise scene with those trained under multi-noise conditions.

		\item[\ding{111}] Can the utilization of the adapter lead to further improvements for SE-based noise-robust ASR system?\\
		Utilizing a SE front-end can significantly improve the performance of ASR systems. 
		The SE front-end enhances features and adapts to the system at the feature level to a certain extent. 
		Hence, while addressing noise-robust ASR challenges, the emphasis is frequently placed on enhancing the performance of the SE front-end. 
		This could result in adapters being relatively uncommon in noise-robust ASR tasks. 
		Hence, in this paper, we investigate whether adapters can further enhance the performance of models adapted at the feature level. 
		
	\end{itemize}
	The adapter is positioned after the encoder layer, shown in Figure~\ref{flowchart}--(b). 
	We perform pretraining on the ASR backend at the outset using a substantial dataset. 
	Following this, we freeze all parameters of the pretrained ASR backend and insert the adapter after the encoder layer. 
	To examine the adapter's impact in unseen noise scenarios, the training noise for the pretrained ASR back-end differs from the noise used for evaluation. 
	The SE-based system is shown in Figure~\ref{flowchart}--(c).

	\vspace{-5pt}

	\section{Experiments}
	\label{experiments}
	\vspace{-5pt}
	\subsection{Experimental Settings}
	\vspace{-5pt}
	The experiments conducted utilized the CHiME-4 dataset\footnote{https://spandh.dcs.shef.ac.uk/chime\_challenge/CHiME4/index.html}.
	It includes four different noise conditions: bus (bus), cafe (caf), pedestrian area (ped), and street junction (str). 
	The audio in the dataset was digitized at a sampling rate of 16 kHz.
	Data from channels 1 to 6 were utilized during the model training phase. 
	The Channel~5 real noisy data from development and evaluation sets were used for testing.

	The Conformer-based ASR backend employed 6 encoder layers. 
	Each encoder layer utilized the relative positional encoding module for positional encoding, encompassing a subsampling rate of 4 facilitated by 2 Conv2d layers. 
	The multi-head attention dimension was set at 512 with 4 attention heads. 
	Position-wise feedforward units numbered 2048, and the activation function used was swish. 
	The input LMFB feature was 80-dimension. 
	The decoder, based on the Transformer architecture \cite{8462506}, comprised 6 decoder layers, each incorporating 512 dimensions for multi-head attention with 4 attention heads. 
	Hidden units for position-wise feedforward were specified as 2048. 
	The dictionary was constructed utilizing transcripts from CHiME-4, WSJ0 \footnote{https://catalog.ldc.upenn.edu/LDC93s6a}, and WSJ1 \footnote{https://catalog.ldc.upenn.edu/LDC94S13A}. 
	The BPE vocabulary consisted of 1014 elements, which included $<blank>$, $<unk>$, and $<sos/eos>$. 
	The ASR backend was pretrained using WSJ0, WSJ1, and Librispeech (960 hours) \cite{7178964}. 
	In the pretraining phase, we synthesized noisy speech by combining the MUSAN dataset \cite{snyder2015musan} with clean speech, randomly selecting signal-to-noise ratios (SNRs) within the range of 0 to 20 dB. 
	Pretraining was conducted 100 epochs with the SpecAug \cite{Park2019}. 
	The final pretrained model was an average of ten well-performing checkpoints on the development set.

	``Bi-LSTM'' and ``DEMUCS'' \cite{Dfossez2020} were chosen as the SE front-end. 
	For ``Bi-LSTM'', the feature used was the magnitude of the spectrogram. 
	The architecture included two Bi-directional LSTM (Bi-LSTM) layers and a fully connected layer. 
	Each Bi-LSTM layer was equipped with 896 hidden nodes. 
	There were 896 hidden nodes in each Bi-LSTM layer. 
	For ``DEMUCS'', we followed the original neural network architecture\footnote{https://github.com/facebookresearch/denoiser}. 
	The SE frontend was pretrained with the CHiME--4 dataset with 200 epochs.

	For the adapter, the input and output dimensions were all 512. 
	During adapter training, all parameters within the pretrained ASR backend remained frozen. 
	We positioned the adapters following the encoder layers. 
	Moreover, both the SE frontend and the adapters were updated simultaneously when utilizing the SE frontend. 
	
	\vspace{-5pt}
	
	\begin{table*}[h]
		\centering
		\caption{The performance of baseline pretrained ASR model.}
		\begin{tabular}{c|l|ccccc|ccccc}
			\toprule[1pt]
			\multirow{2}{*}{\textbf{Exp.}} &
			\multirow{2}{*}{\textbf{System}} &
			\multicolumn{5}{c|}{\textbf{Development Sets}} &
			\multicolumn{5}{c}{\textbf{Evaluation Sets}} \\
			&
			&
			\textbf{bus} &
			\multicolumn{1}{c}{\textbf{str}} &
			\multicolumn{1}{c}{\textbf{ped}} &
			\multicolumn{1}{c}{\textbf{caf}} &
			\multicolumn{1}{c|}{\textbf{avg.}} &
			\textbf{bus} &
			\multicolumn{1}{c}{\textbf{str}} &
			\multicolumn{1}{c}{\textbf{ped}} &
			\multicolumn{1}{c}{\textbf{caf}} &
			\multicolumn{1}{c}{\textbf{avg.}} \\
			\midrule
			\rowcolor[HTML]{EFEFEF} 0 & Pretrained & 22.0 & 15.1 & 11.2 & 16.6 & 16.3 & 30.3 & 16.1 & 23.4 & 26.5 & 24.1 \\
			\bottomrule[1pt]
		\end{tabular}
		\vspace{-10pt}
		\label{pre_trained}
	\end{table*}

	\begin{table*}[h]
		\centering
		\caption{The effect of placing the adapter into different encoder layers (\textbf{trained using the entire CHiME--4 training dataset}).}
		\begin{tabular}{c|cccccc|ccccc|ccccc}
			\toprule[1pt]
			\multirow{2}{*}{\textbf{Exp.}} &
			\multicolumn{6}{c|}{\textbf{Position of Adapter (Encoder)}} &
			\multicolumn{5}{c|}{\textbf{Development Sets}} &
			\multicolumn{5}{c}{\textbf{Evaluation Sets}} \\
			&
			\textbf{E1} &
			\multicolumn{1}{c}{\textbf{E2}} &
			\multicolumn{1}{c}{\textbf{E3}} &
			\multicolumn{1}{c}{\textbf{E4}} &
			\multicolumn{1}{c}{\textbf{E5}} &
			\multicolumn{1}{c|}{\textbf{E6}} &
			\textbf{bus} &
			\multicolumn{1}{c}{\textbf{str}} &
			\multicolumn{1}{c}{\textbf{ped}} &
			\multicolumn{1}{c}{\textbf{caf}} &
			\multicolumn{1}{c|}{\textbf{avg.}} &
			\textbf{bus} &
			\multicolumn{1}{c}{\textbf{str}} &
			\multicolumn{1}{c}{\textbf{ped}} &
			\multicolumn{1}{c}{\textbf{caf}} &
			\multicolumn{1}{c}{\textbf{avg.}} \\
			\midrule
			\rowcolor[HTML]{EFEFEF} 1 & & & & & & \ding{51} &17.1 &10.3 &8.5 &11.1 &11.8 &25.9 &12.1 &17.8 &20.4 &19.1	\\
			2 & & & & & \ding{51} & &15.9 &9.6 &7.9 &9.8 &10.8 &24.4 &11.5 &16.3 &19.1 &17.8 \\		
			\rowcolor[HTML]{EFEFEF} 3 & & & & \ding{51}	& & &14.3 &8.8 &7.2 &9.2 &9.9 &23.1 &10.6 &15.0 &17.1 &16.5 \\
			
			4 & & & \ding{51} & & & &14.4 &8.6 &6.8 &8.9 &9.7 &23.1 &10.8 &15.2 &17.5 &16.7	\\
			
			\rowcolor[HTML]{EFEFEF} 5 & & \ding{51} & & & & &13.5 &8.0 &\textcolor{blue}{\textbf{6.6}} &8.2 &9.1 &21.6 &\textcolor{blue}{\textbf{9.8}} &13.9 &16.2 &15.4	\\
			
			6 & \ding{51} & & & & & &\textcolor{blue}{\textbf{13.4}} &\textcolor{blue}{\textbf{7.8}} &6.7 &\textcolor{blue}{\textbf{8.0}} &\textcolor{blue}{\textbf{9.0}} &\textcolor{blue}{\textbf{20.8}} &10.2 &\textcolor{blue}{\textbf{13.7}} &\textcolor{blue}{\textbf{15.9}} &\textcolor{blue}{\textbf{15.1}}	\\
			
			\midrule
			\rowcolor[HTML]{EFEFEF} 7 & \ding{51} & \ding{51} & & & & &13.0 &7.9 &6.5 &8.1 &8.9 &20.3 &9.9 &13.7 &16.1 &15.0 \\
			8 & \ding{51} & \ding{51} & \ding{51} & & & &13.1 &7.8 &6.5 &8.2 &8.9 &20.4 &10.1 &13.8 &16.1 &15.1	\\
			\rowcolor[HTML]{EFEFEF} 9 & \ding{51} & \ding{51} & \ding{51} & \ding{51} & & &12.9 &7.9 &6.8 &7.9 &8.9 &20.5 &9.6 &13.3 &15.6 &14.7	\\
			10 & \ding{51} & \ding{51} & \ding{51} & \ding{51} & \ding{51} & &12.9 &7.9 &6.7 &8.0 &8.9 &20.4 &9.5 &13.7 &15.6 &14.8	\\
			\rowcolor[HTML]{EFEFEF} 11 & \ding{51} & \ding{51} & \ding{51} & \ding{51} & \ding{51} & \ding{51} &13.1 &7.9 &6.9 &8.1 &9.0 &20.3 &9.6 &13.3 &15.6 &14.7	\\
			\bottomrule[1pt]
		\end{tabular}
		\vspace{-10pt}
		\label{different_layers}
	\end{table*}

	\begin{table*}[h!]
		\centering
		\caption{The effect of different embedding dimensions in adapter (\textbf{trained using the entire CHiME--4 training dataset}).}
		\begin{tabular}{c|c|ccccc|ccccc}
			\toprule[1pt]
			\multicolumn{1}{c|}{\multirow{2}{*}{\textbf{Exp.}}} &
			\multicolumn{1}{c|}{\multirow{2}{*}{\textbf{Emb. Dim.}}} &
			\multicolumn{5}{c|}{\textbf{Development Sets}} &
			\multicolumn{5}{c}{\textbf{Evaluation Sets}} \\
			\multicolumn{1}{c|}{} &
			\multicolumn{1}{c|}{} &
			\multicolumn{1}{c}{\textbf{bus}} &
			\multicolumn{1}{c}{\textbf{str}} &
			\multicolumn{1}{c}{\textbf{ped}} &
			\multicolumn{1}{c}{\textbf{caf}} &
			\multicolumn{1}{c|}{\textbf{avg.}} &
			\multicolumn{1}{c}{\textbf{bus}} &
			\multicolumn{1}{c}{\textbf{str}} &
			\multicolumn{1}{c}{\textbf{ped}} &
			\multicolumn{1}{c}{\textbf{caf}} &
			\multicolumn{1}{c}{\textbf{avg.}} \\
			\midrule
			\rowcolor[HTML]{EFEFEF} 12 &16 & 13.1	&7.9	&6.8	 &8.1 &9.0 &20.6	&9.8	&13.4	&15.6 &14.9  \\
			13 &32 &13.2	&7.8	&6.8	&8.1 &9.0 &20.8			&9.9	&13.4	&15.9 &15.0 \\
			\rowcolor[HTML]{EFEFEF} 11 &64 &13.1 &7.9 &6.9 &8.1 &9.0 &20.3 &9.6 &13.3 &15.6 &14.7 \\
			14 &96 & 13.3		&	8.1		&	6.9	&		7.7 & 9.0 & 20.6		&	9.6		&	13.5		&	15.8 & 14.9 \\
			\rowcolor[HTML]{EFEFEF} 15 &128 & 12.9		&	7.9	&		6.9		&	8.0 & 8.9 & 20.6		&	9.8		&	13.6		&	15.6 & 14.9 \\
			\bottomrule[1pt]
		\end{tabular}
		\vspace{-10pt}
		\label{different_dimension}
	\end{table*}

	\subsection{Effect of the Position of Adapter }
	\vspace{-5pt}
	Table~\ref{different_layers} shows the comparison of placing the adapters into different encoder layers. 
	Compared with the pretrained models in Table~\ref{pre_trained}, inserting the adapter into any encoder layer can bring considerable performance improvement to the system. 
	Through experiments 1--6, it is more effective to insert the adapter in the shallow layer: the performance of the adaptation gradually improves as the number of layers becomes shallower. 
	Some studies \cite{9054675} have revealed that the shallower layers within the ASR model encompass signal-level information, such as speech structure. In contrast, the deeper layers tend to hold abstract information. 
	Therefore, the shallow layer may have more noise-related information, which is why the shallow layer is more effective for adaptation. 
	
	In addition, we also compared the effects of multi-layer adaptation through experiments 7--11. 
	Compared to solely adapting the first encoder layer, incorporating further adaptations in deeper layers did not result in substantial performance enhancements. 
	Considering the abovementioned analysis, leveraging more noise-related information, the self-adaptation at the shallow layer has already achieved satisfactory effect. 
	Attempting to enhance information encoding in the deep layer by reducing noise-related information is challenging, resulting in minimal performance improvements. 
	However, the best performance was achieved by inserting the adapter after all encoder layers on the evaluation sets (experiment 11). 
	Thus, in subsequent experiments, we placed adapters after all encoder layers.

	\vspace{-5pt}

	\subsection{Effect of the Adapter Embedding Dimension}
	\vspace{-5pt}
	Table~\ref{different_dimension} shows the comparison of different embedding dimensions in the adapter. 
	We tried adapters with embedding dimensions of 16, 32, 64, 96, and 128 (experiments 11--15). 
	Despite the wide range of embedding dimensions, the results demonstrate consistent performance. 
	This indicates a degree of robustness in the adapter, as it appears unaffected by the embedding dimension. 
	Based on the results of these models on the evaluation sets, we ultimately chose to utilize the 64-dimensional embedding adapter for the subsequent experiments. 
	
	\vspace{-5pt}

	\begin{table*}[h]
		\centering
		\caption{The effect of different training sets for adapter-based adaptation. ``\textbf{Held}'' represents whether the specific noise conditions were excluded during model training; when utilizing the held-out approach, both the training and testing sets utilize a single noise type condition. ``\textbf{Real}'' represents whether the real noisy data is used during the training process. ``\textbf{Simu.}'' represents whether the simulated noisy data is used during training.
			``\textbf{Utt.}'' represents how many utterances (channels 1 to 6 of the same utterance are considered single utterances) are used during the training process. 
			$\clubsuit$  represents the number of all utterances in the corresponding noisy condition (this is due to the slightly different amounts of simulated data for the four noise conditions). 
			$\bigstar$ represents that 100 sentences are selected from four noise conditions to constitute a training set. 
		}
		\begin{tabular}{c|ccccc|ccccc|ccccc}
			\toprule[1pt]
			\multicolumn{1}{c|}{\multirow{2}{*}{\textbf{Exp.}}} &
			\multicolumn{5}{c|}{\textbf{Training Data}} &
			\multicolumn{5}{c|}{\textbf{Development Sets}} &
			\multicolumn{5}{c}{\textbf{Evaluation Sets}} \\
			\multicolumn{1}{c|}{} &
			\multicolumn{1}{c}{\textbf{Held}} &
			\multicolumn{1}{c}{\textbf{Real}} &
			\multicolumn{1}{c}{\textbf{Utt.}} &
			\multicolumn{1}{c}{\textbf{Simu.}} &
			\multicolumn{1}{c|}{\textbf{Utt.}} &
			\multicolumn{1}{c}{\textbf{bus}} &
			\multicolumn{1}{c}{\textbf{str}} &
			\multicolumn{1}{c}{\textbf{ped}} &
			\multicolumn{1}{c}{\textbf{caf}} &
			\multicolumn{1}{c|}{\textbf{avg.}} &
			\multicolumn{1}{c}{\textbf{bus}} &
			\multicolumn{1}{c}{\textbf{str}} &
			\multicolumn{1}{c}{\textbf{ped}} &
			\multicolumn{1}{c}{\textbf{caf}} &
			\multicolumn{1}{c}{\textbf{avg.}} \\
			\midrule
			\rowcolor[HTML]{EFEFEF} 11 &\ding{55} &\ding{51} &1,600 &\ding{51} &7,138 &13.1 &7.9 &6.9 &8.1 &9.0 &20.3 &9.6 &13.3 &15.6 &14.7  \\
			\rowcolor[HTML]{EFEFEF} 16 &\ding{55} &\ding{51} &1,600 &\ding{55} &- &13.7 &8.2 &7.1 &8.6 &9.4 &22.6 &9.8 &14.1 &16.7 &15.8 \\
			\midrule
			17 &\ding{51} &\ding{51} &400 &\ding{51} &$\clubsuit$ &13.6 &8.2 &7.0 &8.4 &9.3 &21.3 &10.2 &13.6 &15.6 &15.2 \\
			18 &\ding{51} &\ding{51} &400 &\ding{55} &- &14.5 &8.8 &7.2 &9.1 &9.9 &23.2 &10.3 &14.8 &17.8 &16.5 \\
			19&\ding{51} &\ding{55} &- &\ding{51} &400 &15.1 &8.9 &7.5 &9.1 &10.1 &23.6 &10.5 &15.9 &17.4 &16.9 \\
			\midrule
			\rowcolor[HTML]{EFEFEF} 20 &\ding{51} &\ding{51} &200 &\ding{51} &$\clubsuit$ &13.7 &8.1 &7.0 &8.4 &9.3 &21.4 &10.1 &13.9 &15.9 &15.3 \\
			\rowcolor[HTML]{EFEFEF} 21 &\ding{51} &\ding{51} &200 &\ding{55} &- &16.6 &9.6 &7.9 &10.4 &11.1 &25.4 &13.4 &18.1 &21.1 &19.5 \\
			\rowcolor[HTML]{EFEFEF} 22&\ding{51} &\ding{55} &- &\ding{51} &200 &18.1 & 10.0 & 8.9 & 11.7 &12.2 &26.4 &13.6 &20.2 &22.0 &20.5 \\
			\midrule
			23 &\ding{51} &\ding{51} &100 &\ding{51} &$\clubsuit$ &13.9 &8.2 &7.0 &8.5 &9.4 &22.0 &10.2 &13.8 &16.1 &15.5 \\
			24 &\ding{51} &\ding{51} &100 &\ding{55} &- &18.4 &11.2 &8.8 &11.7 &12.5 &27.2 &14.7 &20.3 &23.1 &21.3 \\
			25 &\ding{51} &\ding{55} &- &\ding{51} &100 &20.6 &11.9 &10.6 &13.8 &14.2 & 29.4 &15.2 &22.2 &24.5 &22.8 \\
			\midrule
			\rowcolor[HTML]{EFEFEF} 26 &\ding{55} &\ding{51} &$\bigstar$400  &\ding{55} &- &14.8 &8.7 &7.2 &8.9 &9.9 &23.2 &10.5 &14.7 &17.3 &16.4 \\
			\bottomrule[1pt]
		\end{tabular}
		\vspace{-10pt}
		\label{different_training_sets}
	\end{table*}

	\begin{table*}[h]
		\centering
		\caption{The effect of adapter for different SE-based robust ASR systems (\textbf{trained using the entire CHiME--4 training dataset}).}
		\begin{tabular}{c|l|ccccc|ccccc}
			\toprule[1pt]
			\multirow{2}{*}{\textbf{Exp.}} &
			\multirow{2}{*}{\textbf{System}} &
			\multicolumn{5}{c|}{\textbf{Development Sets}} &
			\multicolumn{5}{c}{\textbf{Evaluation Sets}} \\
			&
			&
			\textbf{bus} &
			\multicolumn{1}{c}{\textbf{str}} &
			\multicolumn{1}{c}{\textbf{ped}} &
			\multicolumn{1}{c}{\textbf{caf}} &
			\multicolumn{1}{c|}{\textbf{avg.}} &
			\textbf{bus} &
			\multicolumn{1}{c}{\textbf{str}} &
			\multicolumn{1}{c}{\textbf{ped}} &
			\multicolumn{1}{c}{\textbf{caf}} &
			\multicolumn{1}{c}{\textbf{avg.}} \\
			\midrule
			\rowcolor[HTML]{EFEFEF} 27 & Bi-LSTM &13.3 &8.1 &6.8 &8.4 &9.2 &21.4 &10.1 &13.5 &16.1 &15.3  \\
			28 & + adapter &12.4 &7.5 &6.7 &7.9 &8.6 &19.7 &9.8 &12.4 &14.1 &14.0 \\
			\rowcolor[HTML]{EFEFEF} 29 & DEMUCS &11.7 &7.7 &6.3 &6.9 &8.2 &19.2 &9.2 &12.8 &14.9 &14.0  \\
			30 & + adapter &10.7 &6.7 &6.4 &6.5 &7.6 &16.8 &8.3 &11.4 &13.2 &12.4 \\
			\bottomrule[1pt]
		\end{tabular}
		\vspace{-10pt}
		\label{front_end}
	\end{table*}

	\subsection{Effect of the Training Data}
	\vspace{-5pt}
	Table~\ref{different_training_sets} shows the comparison of different training sets for adapter training. 
	When comparing experiments 11 and 16, it becomes evident that incorporating simulated data during training leads to better performance for real noisy sets: the relative improvement was 4\% and 7\% for development and evaluation sets, respectively. 
	This trend becomes more pronounced when the quantity of real data diminishes: the relative improvements for experiments 17--18, 20--21, 23--24 were 6\%, 16\%, and 25\% in development sets; and 8\%, 22\%, and 27\% in evaluation sets, respectively. 
	Nevertheless, using simulated data might constrain the model from achieving further improvements in performance with real data. 
	When comparing experiments 18--21 (21--24), it becomes evident that the addition of more real data can lead to substantial performance enhancements, resulting in relative improvements of 11\% (11\%) and 15\% (9\%) for development and evaluation sets, respectively. 
	On the other hand, incorporating the same simulated data did not have any effect in the comparative experiments 17, 20, and 23.

	Real data yields better adaptation performance when using the same amount of data than simulated data. 
	This becomes particularly noticeable, especially when dealing with smaller amounts of data: the relative improvements for experiments 18--19, 21--22, 24--25 were 2\% (development sets), 9\% (development sets), and 12\% (development sets); and 2\% (evaluation sets), 5\% (evaluation sets), and 7\% (evaluation sets), respectively. 
	This could be attributed to the distinct distribution of simulated data compared to real data. 
	Nevertheless, with a substantial volume of simulated data, certain instances might exhibit a distribution comparable to real data, consequently improving the model's performance on the real test sets.

	Furthermore, we investigated how multi-condition training influences the adapter's effectiveness. 
	The performance of experiments 18 and 26 were the same. 
	This could be due to shared noises (because each noise condition is composed of multiple noises) among the four noise scenes in the CHiME--4 dataset. 
	As a result, the noise is only partially unseen, thus partially limiting the potential effects of multi-condition training. 
	It also serves as an inspiration that incorporating similar noisy real data as augmented data can result in substantial performance improvements (compare experiments 24 and 26).

	\vspace{-5pt}

	\subsection{Effect of the Adapter for SE-based robust ASR}
	\vspace{-5pt}
	Table~\ref{front_end} shows the adapter's performance for different SE-based robust ASR systems. 
	Experiments 27 and 29 significantly improved the performance from the pretrained model (experiment 0) when the SE front-end was used. 
	Incorporating adapters within the SE-based robust ASR system further improved recognition performance. 
	While feature enhancement of the SE front-end has been achieved, significant benefits still arise from adaptation at the backend. 
	This is because the SE frontend might introduce information loss or distortion and the adapter plays a role in mitigating these issues. 
	
	\vspace{-5pt}

	\section{Conclusions and Future Works}
	\vspace{-5pt}
	In this paper, we explored the effect of the adapter on noise-robust ASR. 
	We conducted a comprehensive exploration from various perspectives, including the optimal insertion position for the adapter, the quantity and type of data used for adapter training, and the synergy of the adapter with the SE. 
	We conducted experiments using the CHiME--4 dataset. 
	The experimental results demonstrate that incorporating adapters in the shallow layer yields more effectness compared to the deep layer. 
	Furthermore, the number of embedding nodes in the adapter does not significantly impact the adaptation process. 
	Moreover, the training dataset plays a vital role in adapter training: 
	When considering the same data amount, using real data shows more effective than simulated data but adding simulated data can enhance the performance of the real test sets. 
	Combining adapters with the SE frontend leads to further performance improvement. 
	In the future, we will study on designing the adapter structure better suited for robust ASR.

	\clearpage
	\footnotesize
	\bibliographystyle{IEEEbib}
	\bibliography{refs}
	
\end{document}